\documentclass[
  aip,
  amsmath,amssymb,
  reprint
]{revtex4-1}

\usepackage[utf8]{inputenc}
\usepackage[T1]{fontenc}
\usepackage{mathptmx}
\usepackage{graphicx}
\usepackage{dcolumn}
\usepackage{bm}

\usepackage{etoolbox}
\makeatletter
\def\@email#1#2{%
 \endgroup
 \patchcmd{\titleblock@produce}
  {\frontmatter@RRAPformat}
  {\frontmatter@RRAPformat{\produce@RRAP{*#1\href{mailto:#2}{#2}}}\frontmatter@RRAPformat}
  {}{}
}%
\makeatother

\begin{document}

\preprint{AIP/ErYIG-PLD}

\title{Magnetic and structural properties of epitaxial Er-substituted yttrium iron garnet films grown by pulsed laser deposition}

\author{Luk{\'{a}}{\v{s}} Flaj{\v{s}}man}
\affiliation{Department of Applied Physics, Aalto University, P.O. Box 15100, FI-00076 Aalto, Finland}
\email{lukas.flajsman@aalto.fi}

\author{Lars Peeters}
\affiliation{Department of Applied Physics, Aalto University, P.O. Box 15100, FI-00076 Aalto, Finland}

\author{Armi Kosunen}
\affiliation{Department of Applied Physics, Aalto University, P.O. Box 15100, FI-00076 Aalto, Finland}


\author{Lide Yao}
\affiliation{OtaNano-Nanomicroscopy Center, Aalto University, P.O. Box 15100, FI-00076 Aalto, Finland}

\author{Ionela Vrejoiu}
\affiliation{II. Physikalisches Institut, University of Cologne, Z{\"u}lpicher Str.\ 77, 50937 K{\"o}ln, Germany}

\author{Sebastiaan van Dijken}
\affiliation{Department of Applied Physics, Aalto University, P.O. Box 15100, FI-00076 Aalto, Finland}
\email{sebastiaan.van.dijken@aalto.fi}

\date{\today}


\begin{abstract}
Er-substituted yttrium iron garnet (Er:YIG) holds the potential of combining the low magnetic damping of YIG with the telecom-band optical transitions of $\text{Er}^{3+}$ ions, making it a suitable material for hybrid optomagnonic devices and microwave-to-optical quantum transduction. We report the epitaxial growth of $\text{Er}_{x}\text{Y}_{3-x}\text{Fe}_{5}\text{O}_{12}$ films with $x=0.008-0.20$ on (111)-oriented gadolinium gallium garnet (GGG) substrates using pulsed laser deposition. X-ray diffraction, reciprocal space mapping, and scanning transmission electron microscopy confirm single-phase, fully coherent growth with atomically sharp interfaces across the entire substitution range. Magnetometry reveals a gradual decrease in saturation magnetization with increasing Er content, consistent with antiparallel coupling between Er$^{3+}$ spins and the net Fe$^{3+}$ moments, along with the emergence of an in-plane uniaxial magnetic anisotropy. The ferromagnetic resonance broadens with Er concentration due to increased Gilbert damping and inhomogeneous linewidth broadening. Films with low Er content ($x=0.008$), most relevant for optomagnonic applications, retain nearly isotropic magnetization and exhibit a damping parameter only slightly higher than that of undoped YIG. These results identify growth and substitution conditions that preserve YIG's low-loss magnetic properties while introducing optical functionality, establishing Er:YIG as a viable platform for hybrid quantum magnonics and microwave-to-optical transduction.
\end{abstract}

\maketitle

\section{Introduction}
Quantum magnonics explores the use of collective spin excitations in magnetic materials, magnons, as carriers and mediators of quantum information.\cite{Yuan:2022} Magnons can coherently couple to microwave photons, optical photons, and phonons, providing hybrid interfaces that connect otherwise distinct quantum systems.\cite{Lachance-Quirion:2019,Li:2020,Rameshti:2022} Among these interactions, coherent coupling between magnons and optical photons is particularly relevant for microwave-to-optical quantum transduction, which is essential for linking superconducting quantum processors with long-distance optical communication networks. 

Efficient optomagnonic coupling requires materials that combine low magnetic damping, high spin density, low optical loss, and compatibility with high-Q optical cavities. YIG satisfies these requirements and is the benchmark material for optomagnonic research. It has enabled demonstrations of coherent magnon-optical photon coupling in several geometries, including whispering-gallery-mode spheres,\cite{Osada:2016,Zhang:2016,Haigh:2016} ridge waveguide resonators,\cite{Zhu:2020} and Fabry–Pérot microcavities.\cite{Haigh:2021} In these systems, a magnon mode couples to two orthogonally polarized optical modes (TE and TM) through triple-resonance Brillouin light scattering, enabling frequency transduction between microwave and optical domains. However, the overall transduction efficiency remains low, typically around $10^{-8}$ or less, due to the weak magneto-optical interaction and limited spatial overlap between the magnonic and optical modes.

To address these limitations, new strategies need to focus on integrating magnonic and optical functionalities within a single material platform. One approach involves doping low-damping magnetic materials with rare-earth ions to introduce narrow optical transitions.\cite{Puel:2025} Although pure YIG lacks strong optical emission lines, substituting rare-earth ions into its lattice generates sharp optical transitions that enable coherent interactions among magnons, photons, and rare-earth ion spins. Among possible dopants, Er is particularly advantageous. Er$^{3+}$ ions exhibit narrow intra-4$f$ transitions at 1.5~\textmu m (within the telecom C-band), maintain good optical coherence up to 8~K,\cite{Gritsch:2022} show high chemical compatibility with YIG,\cite{Cho:2022} and possess large magnetic moments (9.6~\textmu$_\mathrm{B}$/Er) that can couple to GHz-frequency magnons. Moreover, Er$^{3+}$ ions possess long spin relaxation times, reaching up to 0.1~s at cryogenic temperatures,\cite{Baldit:2010,Böttger:2009} suggesting that Er:YIG can act as a hybrid medium supporting coherent magnon–photon–spin interactions at the interfaces between microwave and optical quantum systems.

Previous studies on Er:YIG bulk materials, including powders\cite{Cho:2022} and single crystals grown using a two-step floating-zone method,\cite{Cho:2025} have shown that Er$^{3+}$ ions predominantly replace Y$^{3+}$ ions at the dodecahedral sites of the YIG lattice, while preserving the ferrimagnetic ordering of Fe$^{3+}$ ions on the tetrahedral and octahedral sublattices. In this work, we report the epitaxial growth of single-crystal Er:YIG thin films on GGG substrates with varying Er content ($x=0.008-0.20$). We identify systematic trends in lattice-cell volume, saturation magnetization, and substrate-induced magnetic anisotropy as functions of Er-Y substitution and growth temperature. Notably, we demonstrate that magnetic damping remains low at small Er substitution levels, a regime particularly relevant for hybrid optomagnonic systems and microwave-to-optical quantum transduction.      

\section{Experimental details}
Er:YIG films were grown on (111)-oriented GGG substrates by pulsed laser deposition (PLD). Before deposition, the substrates were ultrasonically cleaned in acetone and isopropanol, then annealed \textit{ex situ} at $1000~^{\circ}\mathrm{C}$ for four hours in 1~bar of flowing oxygen to ensure clean, well-ordered surfaces. During PLD, a KrF excimer laser ($\lambda = 248$~nm) operated at 5~Hz with a fluence of 2.5~J.cm$^{-2}$ was focused onto stoichiometric, polycrystalline Er:YIG targets. The target-substrate distance was fixed at 55~mm. To investigate the influence of Er content on the structural and magnetic properties, four target compositions were used:
Er$_{0.008}$Y$_{2.992}$Fe$_5$O$_{12}$ (Er$_{0.008}$:YIG), Er$_{0.05}$Y$_{2.95}$Fe$_5$O$_{12}$ (Er$_{0.05}$:YIG), Er$_{0.10}$Y$_{2.90}$Fe$_5$O$_{12}$ (Er$_{0.10}$:YIG), and Er$_{0.20}$Y$_{2.80}$Fe$_5$O$_{12}$ (Er$_{0.20}$:YIG). Film growth was optimized by adjusting both the substrate temperature and oxygen pressure. Smooth, epitaxial films with sharp Er:YIG/GGG interfaces were obtained for substrate temperatures between $750~^{\circ}\mathrm{C}$ and $850~^{\circ}\mathrm{C}$ at an oxygen pressure of 0.025 mbar. After deposition, the films were cooled at $4~^{\circ}\mathrm{C}\,\mathrm{min}^{-1}$ in the same oxygen atmosphere. Each film was grown using 9000 laser pulses, with the resulting thickness depending on both the Er content and growth temperature (see Table~\ref{tab:thickness}). The film with the lowest Er content (Er$_{0.008}$:YIG) was prepared from a target supplied by a different manufacturer, whose higher density led to a reduced film thickness under otherwise identical PLD conditions.  

\begin{table}[t]
\caption{
Thicknesses of Er$_x$Y$_{3-x}$Fe$_5$O$_{12}$ thin films grown on GGG (111) substrates for different Er contents ($x$) and growth temperatures ($T_\mathrm{g}$). The thicknesses were determined from X-ray reflectivity measurements. All films were deposited under an oxygen pressure of 0.025 mbar using 9000 laser pulses.
}
\label{tab:thickness}
\begin{ruledtabular}
\begin{tabular}{cccc}
\textrm{Er content $x$} &
\textrm{$T_\mathrm{g}=750~^{\circ}\mathrm{C}$} &
\textrm{$T_\mathrm{g}=800~^{\circ}\mathrm{C}$} &
\textrm{$T_\mathrm{g}=850~^{\circ}\mathrm{C}$} \\
\colrule
0.008 & - & - & 65.7~nm \\
0.05 & 75.6~nm & 77.8~nm & 83.7~nm \\
0.10 & 78.2~nm & 74.5~nm & 73.5~nm \\
0.20 & 70.9~nm & 70.4~nm & 72.4~nm \\
\end{tabular}
\end{ruledtabular}
\end{table}

\section{Results and Discussion}

\subsection{Structural Properties}

\begin{figure}[t]
\includegraphics[width=\columnwidth]{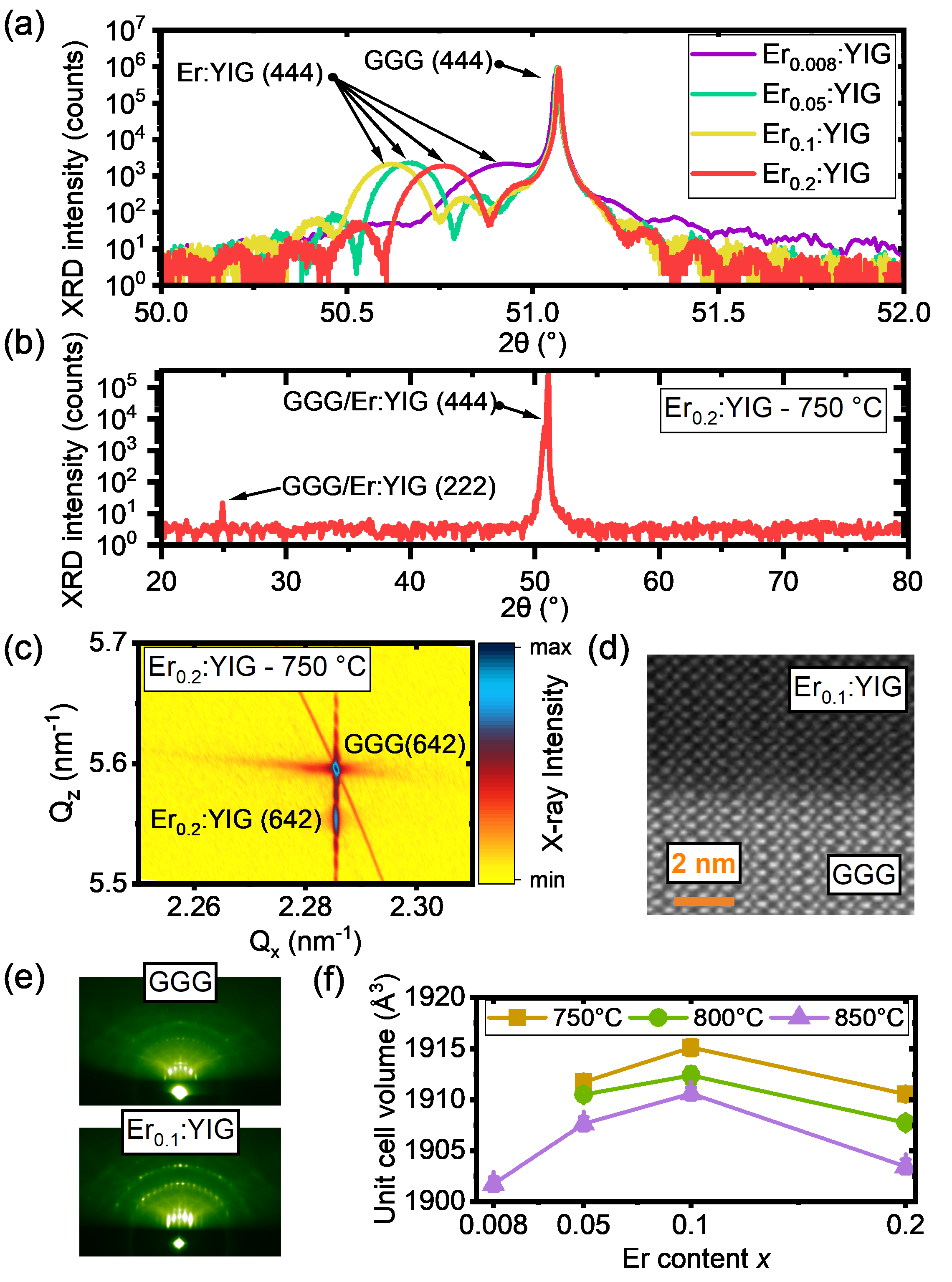}
\caption{Structural characterization of Er:YIG films on GGG (111). (a) XRD $\theta$–$2\theta$ scans for different Er contents ($T_\mathrm{g}=750~^{\circ}\mathrm{C}$). (b) XRD $\theta$–$2\theta$ scan of the Er$_{0.20}$:YIG/GGG sample measured over a larger $2\theta$ range. (c) RSM of the (642) reflection for the Er$_{0.20}$:YIG/GGG sample. (d) STEM image of the Er$_{0.10}$:YIG/GGG interface, imaged along the $[11\bar{2}]$ zone axis. (e) RHEED patterns recorded in the PLD chamber before (top) and after (bottom) Er$_{0.10}$:YIG thin-film growth. (f) Unit cell volumes of all Er:YIG films, extracted from in-plane and out-of-plane lattice constants obtained via XRD (panel (a)) and RSM (panel (c)) measurements, plotted for varying Er contents and growth temperatures.}
\label{fig:fig1}
\end{figure}

Figure~\ref{fig:fig1} summarizes the structural characterization of Er:YIG films with varying Er content. X-ray diffraction (XRD) $\theta$–$2\theta$ scans (Fig.~\ref{fig:fig1}(a)) confirm high-quality epitaxial growth of Er:YIG (111) on GGG (111) substrates for all compositions. The presence of clear Laue fringes around the Er:YIG (444) reflection indicates smooth film surfaces and uniform thickness. The (444) film peak shifts progressively to lower diffraction angles up to $x=0.10$, then moves back to higher angles for $x=0.20$, corresponding to an initial expansion followed by a contraction of the out-of-plane lattice parameter. No secondary phases are detected, as evidenced by the absence of additional diffraction peaks in the wide-range $2\theta=15-80^{\circ}$ scan of the Er$_{0.20}$:YIG/GGG sample (Fig.~\ref{fig:fig1}(b)). Reciprocal space maps (RSMs, Fig.~\ref{fig:fig1}(c)) show identical $Q_x$ values for the films and substrates, confirming coherent in-plane lattice matching and fully strained growth. Scanning transmission electron microscopy (STEM) analysis (Fig.~\ref{fig:fig1}(d)) reveals atomically sharp film-substrate interfaces. Reflection high-energy electron diffraction (RHEED) patterns recorded immediately after growth display sharp streaks and well-defined Kikuchi lines (Fig.~\ref{fig:fig1}(e)), further confirming the high crystalline quality of the films. 

The unit cell volumes were determined from the in-plane lattice parameters obtained by RSM and the out-of-plane parameters from XRD (Fig.~\ref{fig:fig1}(f)). Substitution of Y by Er initially expands the lattice up to $x=0.10$, consistent with the larger ionic radius of Er.\cite{Cho:2022,Sekijima:1999} At higher substitution levels, the lattice contracts, although the films remain epitaxial and single crystalline. Incorporation of rare-earth ions into YIG can introduce structural modifications such as secondary phases, antisite defects, Fe and O vacancies, or other point defects.\cite{Sharma:2018,Tan:2020,Su:2021} Since no secondary phases are observed (Fig.~\ref{fig:fig1}(b)), excess Er is likely accommodated within the garnet lattice through antisite defects (Er occupying Fe sites) or Fe vacancy formation. As rare-earth/Fe antisite defects typically expand the lattice,\cite{Su:2021} the observed contraction at $x=0.20$ suggests that Fe vacancies dominate at higher Er concentrations. In YIG, O ions surrounding an Fe vacancy tend to relax inward, while charge neutrality is maintained by Fe$^{3+}$$-$Fe$^{2+}$ reduction or oxygen vacancy formation. Together, these mechanisms can lead to a net lattice contraction. The formation of Fe vacancies upon Er substitution may also explain the emergence of an in-plane uniaxial magnetic anisotropy,\cite{Manuilov:2009} as discussed in the next section. The absence of significant peak broadening in the XRD data (Fig.~\ref{fig:fig1}(a)) indicates that point defect formation does not substantially degrade the films' long-range crystallinity. Finally, the influence of growth temperature on the lattice parameter is relatively small (Fig.~\ref{fig:fig1}(f)). For all Er substitution levels, increasing the growth temperature slightly reduces the unit cell volume, likely due to enhanced vacancy formation at higher temperatures. 

\subsection{Magnetic Characterization}
\begin{figure}[t]
\includegraphics[width=\columnwidth]{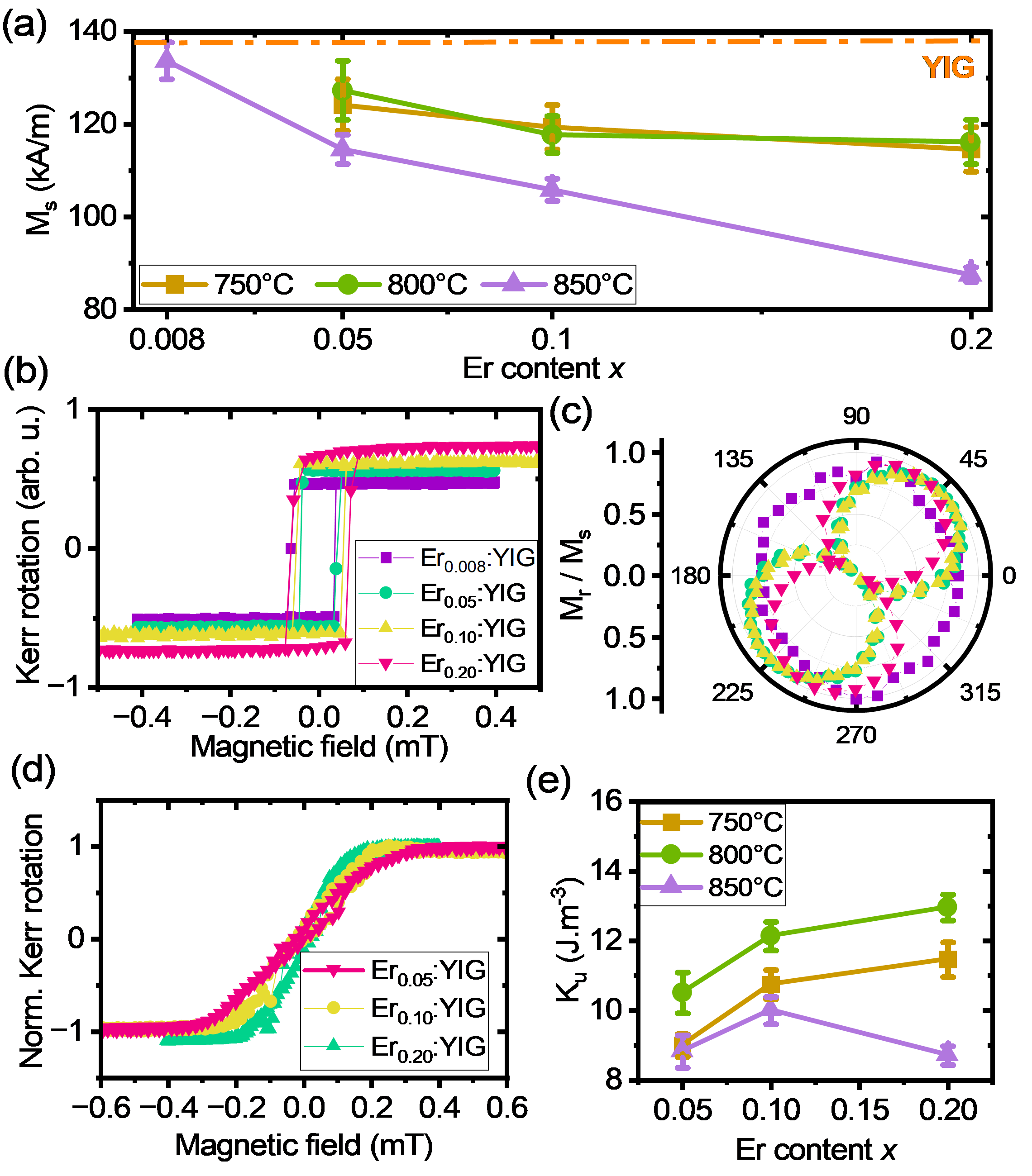}
\caption{(a) Saturation magnetization ($M_\mathrm{s}$) of the Er:YIG films as a function of Er content for samples grown at different substrate temperatures. The dashed-dotted line indicates the $M_\mathrm{s}$ of an undoped YIG reference film. (b) Longitudinal MOKE hysteresis loops recorded at $\lambda=405$~nm with the in-plane magnetic field applied along the easy axis. (c) Polar plots of the remanent magnetization, normalized to $M_\mathrm{s}$, versus the in-plane field angle $\phi$ ($\phi=0^{\circ}$ corresponds to the $[\bar{1}10]$ crystallographic direction). (d) Longitudinal MOKE hysteresis loops measured with the in-plane field applied along the hard axis. Panels (b)-(d) correspond to films grown at $T_\mathrm{g}=800~^{\circ}\mathrm{C}$. (e) Uniaxial magnetic anisotropy constant ($K_\mathrm{u}$) of the Er:YIG films, extracted from the slope of hard-axis hysteresis loops.}
\label{fig:fig2}
\end{figure}

Figure~\ref{fig:fig2}(a) shows the saturation magnetization ($M_\mathrm{s}$) of the Er:YIG films, measured by vibrating sample magnetometry (VSM) for $x=0.05-0.20$ and by vector network analyzer ferromagnetic resonance (VNA-FMR) spectroscopy for $x=0.008$.\cite{Qin:2018} The $M_\mathrm{s}$ decreases systematically with increasing Er content, consistent with antiparallel coupling between Er$^{3+}$ spins at the dodecahedral sites and the net magnetic moment of Fe$^{3+}$ ions on the tetrahedral and octahedral sublattices.\cite{Sekijima:1999} Films grown at $850~^{\circ}\mathrm{C}$ exhibit a further reduction in $M_\mathrm{s}$, which may result from Ga or Gd interdiffusion from the GGG substrate,\cite{Mitra:2017,Cooper:2017,Suturin:2018} or from an increased concentration of Fe or O vacancies.\cite{Tan:2020,Dumont:2007,Noun:2010} 

Longitudinal magneto-optical Kerr effect (MOKE) hysteresis loops measured with the in-plane field applied along the magnetic easy axis (Fig.~\ref{fig:fig2}(b)) exhibit square shapes with low switching fields ($\sim$0.05 mT), indicating efficient domain nucleation and fast domain-wall propagation. The Kerr rotation amplitude increases with Er content, suggesting an enhanced magneto-optical response that compensates for the reduction in $M_\mathrm{s}$ (Fig.~\ref{fig:fig2}(a)). The lightly doped Er$_{0.008}$:YIG film shows no measurable in-plane magnetic anisotropy (Fig.~\ref{fig:fig2}(c)), consistent with undoped YIG (111) films.\cite{Kuznetsov:2025} At higher Er substitution levels, however, an in-plane uniaxial anisotropy emerges, with the magnetic easy axis oriented approximately $50^\circ$ from the $[\bar{1}10]$ crystallographic direction. The uniaxial anisotropy constant ($K_\mathrm{u}$) was estimated by fitting the slope of normalized hard-axis MOKE hysteresis curves (Fig.~\ref{fig:fig2}(d)), following the method of Ref.~\citenum{Weber:1997}. The extracted $K_\mathrm{u}$ values, summarized in Fig.~\ref{fig:fig2}(e), increases with Er content, except for the Er$_{0.20}$:YIG grown at $T_\mathrm{g}=850~^{\circ}\mathrm{C}$, which shows a markedly reduced $M_\mathrm{s}$. 

In-plane uniaxial magnetic anisotropy has also been reported in Ce:YIG\cite{Kehlberger:2015,Hyun:2025} and Fe-deficient YIG films\cite{Manuilov:2009} grown on GGG (111) substrates. In the present Er:YIG films, the progressive increase in Fe vacancies with Er substitution likely contributes to the observed uniaxial anisotropy, the reduction in $M_\mathrm{s}$, and the slight lattice contraction of the Er$_{0.20}$:YIG film.   

\subsection{Ferromagnetic Resonance and Magnetic Damping}
\begin{figure}[t]
\includegraphics[width=\columnwidth]{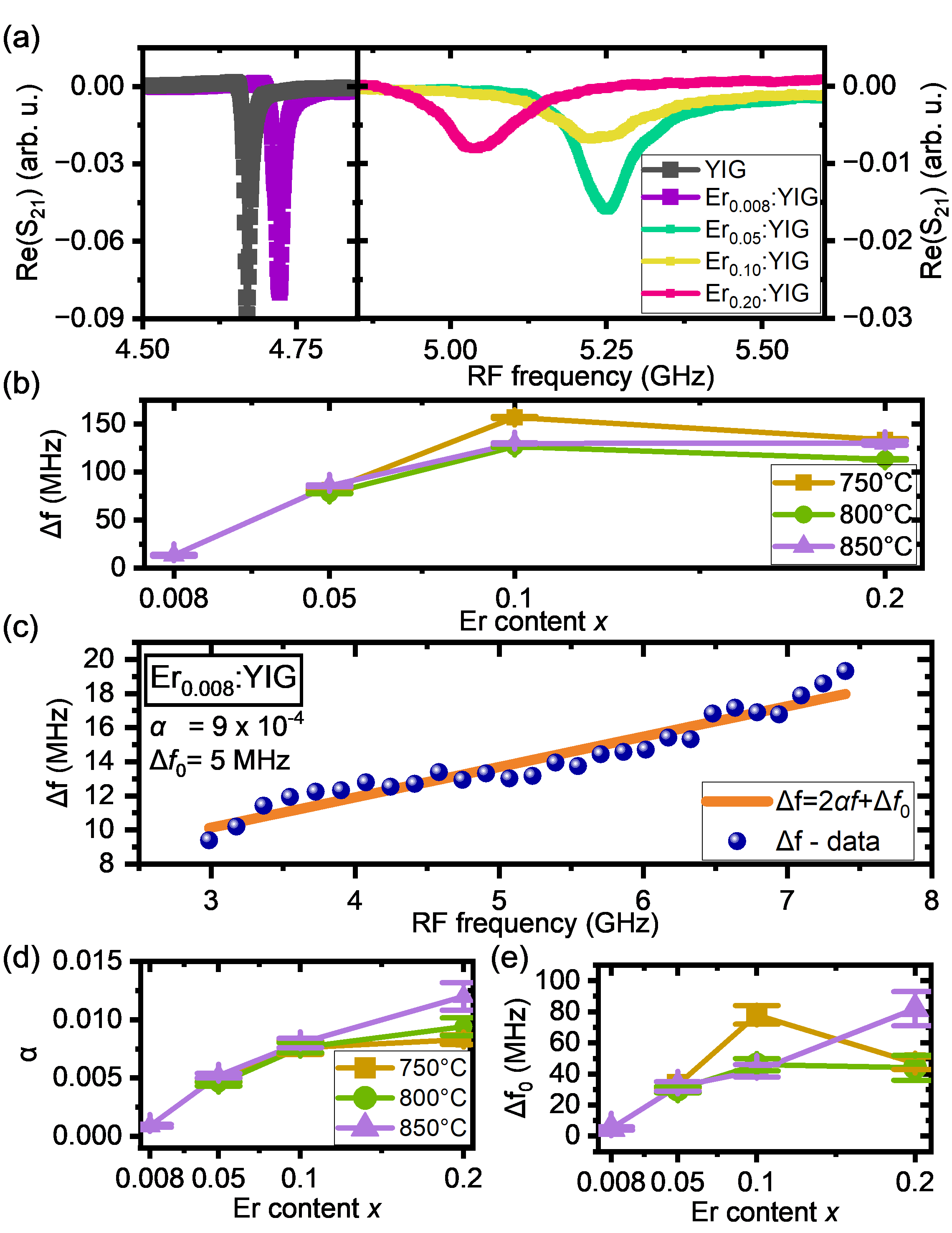}
\caption{(a) VNA-FMR spectra of Er:YIG films and an undoped YIG film measured under an in-plane magnetic field of 100 mT. (b) FMR linewidths of all Er:YIG films measured at 5 GHz. (c) Frequency dependence of the FMR linewidth for the Er$_{0.008}$:YIG film. The linear fit yields a Gilbert damping parameter $\alpha=9\times{10^{-4}}$ and an inhomogeneous linewidth broadening $\Delta{f}_0=5$ MHz. (d) Gilbert damping parameter and (e) inhomogeneous linewidth broadening as a function of Er content.}
\label{fig:fig3}
\end{figure}

The dynamic magnetic properties of the Er:YIG films were characterized using VNA-FMR spectroscopy. During the measurements, the samples were placed face-down on a coplanar waveguide inside a quadrupole electromagnet probe station, with the external magnetic field applied along the in-plane easy axis. FMR spectra were obtained by measuring the S$_{21}$ scattering parameter as a function of frequency under different magnetic bias fields. Figure~\ref{fig:fig3}(a) shows representative FMR spectra measured at 100 mT. The Er$_{0.008}$:YIG film exhibits a clear resonance with a full width at half maximum (FWHM) linewidth of 13.2 MHz, slightly larger than that of the undoped YIG film (9.1 MHz). Increasing the Er content reduces the resonance amplitude and leads to significant linewidth broadening. 

The shifts in resonance frequency observed in Fig.~\ref{fig:fig3}(a) reflect the combined effects of changes in $M_\mathrm{s}$ (Fig.~\ref{fig:fig2}(a)) and $K_\mathrm{u}$ (Fig~\ref{fig:fig2}(e)). The Er$_{0.008}$:YIG film shows a lower resonance frequency despite its higher $M_\mathrm{s}$ due to negligible anisotropy, whereas films with higher Er content exhibit higher resonance frequencies resulting from increased anisotropy. For these higher-substitution films, the gradual frequency downshift with increasing Er content is primarily due to the reduction in $M_\mathrm{s}$.     

Figure~\ref{fig:fig3}(b) summarizes the FMR linewidths of all Er:YIG films measured at 5 GHz. The linewidth increases with Er content, from about 13 MHz for Er$_{0.008}$:YIG to approximately 80 MHz for Er$_{0.05}$:YIG, and up to $120-160$ MHz for Er$_{0.10}$:YIG and Er$_{0.20}$:YIG. To distinguish between inhomogeneous linewidth broadening ($\Delta{f}_0$), arising from spatial variations in magnetic or structural properties, and intrinsic Gilbert damping ($\alpha$), the FMR linewidth was analyzed as a function of frequency (see Fig.~\ref{fig:fig3}(c) for Er$_{0.008}$:YIG). Linear fits to the experimental data using $\Delta{f}=2\alpha{f}+\Delta{f}_0$ yielded values for both the Gilbert damping parameter and inhomogeneous linewidth broadening for all films (Figs.~\ref{fig:fig3}(d) and \ref{fig:fig3}(e)). The Gilbert damping increases monotonically with Er content, consistent with trends in other rare-earth-doped YIG systems.\cite{Sekijima:1999,Dillon:1962} This increase is primarily attributed to enhanced spin-lattice relaxation mediated by the 4$f$ electrons of the rare-earth ions. The rise in inhomogeneous linewidth broadening for higher Er content indicates that different regions of the film resonate at slightly different frequencies, likely due to magnetic or structural inhomogeneities, similar to observations in Bi:YIG.\cite{Das:2023} 

For optomagnonic and microwave-to-optical quantum transduction applications, films with low Er substitution are most relevant. At higher Er concentrations, photoluminescence self-quenching becomes significant, resulting in a reduced Er$^{3+}$ lifetime.\cite{Auzel:2003,Jaba:2009} Our results demonstrate the successful epitaxial growth of single-crystal Er:YIG films over a wide range of substitution levels and identify a narrower regime ($x\lesssim0.01$) suitable for applications requiring both efficient photoluminescence within the telecom C-band and low-loss magnetization dynamics.  

\section{Conclusions}
We demonstrated the epitaxial growth of Er:YIG films on GGG (111) substrates by PLD over a range of growth temperatures. The films are single crystalline, exhibit smooth surfaces, and form a sharp interface with the substrate. Increasing the Er content systematically alters the magnetic properties: the saturation magnetization decreases, an in-plane uniaxial magnetic anisotropy develops, and both the Gilbert damping parameter and inhomogeneous FMR linewidth broadening increase. At low substitution levels, these changes remain modest, allowing the coexistence of optical functionality in the telecom C-band and coherent magnetization dynamics at gigahertz frequencies.

\begin{acknowledgments}
This work was supported by the Research Council of Finland (Grant Nos. 357211 and 359125). L.P. acknowledges financial support from Institute Q and the Wihuri Foundation. We acknowledge the provision of facilities by Aalto University at OtaNano-Nanomicroscopy Center (Aalto-NMC). 
\end{acknowledgments}

\section*{Data Availability}
The data that support the findings of this study are available from the corresponding author upon reasonable request.

\bibliographystyle{aipnum4-1}
\bibliography{aipsampCorr} 

\end{document}